 \documentclass[final,5p,times,twocolumn]{elsarticle}

\usepackage{epsfig}

\usepackage{amssymb}
\journal{Physics Letters B}
\usepackage{hyperref}
\journal{Journal of \LaTeX\ Templates}

\begin{document}

\begin{frontmatter}

\title{Strange stars with a mirror-dark-matter core confronting with the observations of compact stars}

%可能的标题：
%Effects of mirror dark matter on strange stars and constraints
%from the observations of compact stars


%%author{Shu-Hua Yang \corref{ysh@phy.ccnu.edu.cn},
%%Xiao-Ping Zheng, Chun-Mei Pi}

\author{Shu-Hua Yang\corref{cor1}$^{a}$}
\ead{ysh@mail.ccnu.edu.cn}
\author{ Chun-Mei Pi\corref{cor1}$^{b}$}
\author{Xiao-Ping Zheng$^{a,c}$}
\address{$^{a}$ Institute of Astrophysics, Central China Normal University, Wuhan 430079, China}
\address{$^{b}$ School of Physics and Mechanical \& Electrical Engineering, Hubei University of Education, Wuhan 430205, China}
\address{$^{c}$ Department of Astronomy, School of physics, Huazhong University of Science and Technology, Wuhan 430074, China}

\begin{abstract}
We investigate the structure and the tidal deformability of strange stars (SSs) with a mirror-dark-matter (MDM) core for the standard MIT bag model. We find that to explain the observations of PSR J0740+6620, PSR J0030+0451 and GW170817 simultaneously, SSs in GW170817 should have a MDM core although it is unnecessary for PSR J0740+6620 and PSR J0030+0451 to contain a MDM core. Our study leads to the result that for the standard MIT bag model, the observations of compact stars mentioned above confirm the existence of a dark-matter core inside SSs.

\end{abstract}


\end{frontmatter}

%\linenumbers

%%%%%%%%%%%%%%%%%%%%%%%%%%%%%%%%%%%%%%%%%%%%%%%%%%%%%%%%%%%%%%%%%%%%%%%%%%%%%
\section{Introduction}

As hypothesized by Itoh \citep{Itoh1970}, Bodmer \citep{Bodmer1971}, Witten \citep{Witten1984}, and Terazawa \citep{Terazawa1989}, strange quark matter (SQM) consisting of up ($u$), down ($d$) and strange ($s$) quarks and electrons may be the true ground state of baryonic matter. According
to this hypothesis, compact stars made entirely of SQM, referred to as SSs, ought to exist in the universe \cite{Farhi1984,Alcock1986,Haensel1986,Alcock1988,Madsen1999,Weber2005}.

Yang et al. \cite{Yang2020,Yang2021} found that both for the standard MIT bag model and for the density dependent quark mass model, the existence of SSs is ruled out by the mass of PSR J0740+6620 ($2.14_{-0.09}^{+0.10}\, M_{\odot}$ for a 68.3\% credibility interval; $2.14_{-0.18}^{+0.20}\, M_{\odot}$ for a 95.4\% credibility interval) \cite{Cromartie2020} and the dimensionless tidal deformability of a $1.4\, M_{\odot}$ star of
GW170817 ($\Lambda(1.4)=190 _{-120}^{+390}$) \cite{Abbott2017,Abbott2018}. However, Yang et al. \cite{Yang2020,Yang2021} demonstrated that if non-Newtonian gravity effects are considered, SSs can exist for certain ranges of the values of the non-Newtonian gravity parameter. In this Letter, instead of employing the non-Newtonian gravity effects, we propose an alternative explanation to the above mentioned astrophysical observations which supposes that SSs have a mirror-dark-matter (MDM) core.

Compact stars might contain a dark-matter core made of self-interacting dark matter \cite{Spergel2000,Tulin2018, Bertone2018}.
Neutron stars (NSs) and SSs with a dark-matter core have been studied extensively \citep[e.g.,][]{Leung2011,Li2012,Li2012a,Xiang2014,Mukhopadhyay2017,Ellis2018,Ellis2018a,Rezaei2018,Wang2019,
Deliyergiyev2019,Bezares2019,Ivanytskyi2020,Bauswein2020,Das2020a,Ciancarella2020,Mukhopadhyay2016,Panotopoulos2017}.
Especially, Neutron stars with a MDM core have been studied \cite{Sandin2009,Ciarcelluti2011}. Ciarcelluti and Sandin \cite{Ciarcelluti2011} found that the discrepancy between the mass and radius data of EXO 0748-676 \cite{Ozel2006} and the
group 4U 1608-52, 4U 1820-30 and EXO 1745-248 can be interpreted as the signature of a dark-matter core inside them. One key point of their work is that a MDM core inside NSs leads to an apparent softening of the EOS, and the relative amount of MDM in NSs strongly effect the mass-radius relationship of the star. Another key point is that the equilibrium sequence of NSs is non-unique and history dependent, because the relative amount of MDM trapped in each NS could be different, which depends on the individual history, starting from the formation of the progenitor star and continuing through the evolutionary phases until present age.

We will study the structure and the tidal deformability of SSs with a MDM core and explain the observations of PSR J0740+6620, GW170817 and the mass and radius of PSRJ0030+0451 observed by NICER \cite{Riley2019,Miller2019} simultaneously. This Letter is organized as follows: In Sec.\ \ref{EOS}, we briefly review the EOS of SQM and MDM. In Sec.\ \ref{struc}, we present the theoretical framework the structure and the tidal deformability of SSs with a MDM core. In Sec.\ \ref{results}, numerical results and discussions are presented. Finally, the summary and conclusions are given in Sec.\ \ref{Summary}.

\section{EOS of SQM and MDM}\label{EOS}

Following Yang et al. \cite{Yang2020}, we briefly review the phenomenological model for the EOS employed in this paper, namely, the standard bag model \cite{Farhi1984,Alcock1986,Haensel1986,Weber2005}. In that model, $u$ and $d$ quarks are treated as massless particles but
$s$ quarks have a finite mass, $m_s$.  First-order perturbative corrections in the strong interaction coupling constant $\alpha_{S}$
are taken into account. The thermodynamic potential for the $u$, $d$ and $s$ quarks, and for the electrons ($\Omega_{i}$ with $i=u,d,s,e$) are the same as Refs.\cite{Alcock1986,Yang2020} with the renomormalization constant $\sigma=300$ MeV, and the current mass of $s$ quark is taken as $m_s=93$ MeV \citep{Zyla2020}. The number density of each species is given by
\begin{equation}
n_{i}=-\frac{\partial\Omega_{i}}{\partial\mu_{i}},
\end{equation}
where $\mu_{i}$ ($i=u,d,s,e$) are the chemical potentials. For SQM, chemical equilibrium is maintained by the weak-interaction, which leads for the chemical potentials to the following conditions,
\begin{eqnarray}
\mu_{d} &=&  \mu_{s} , \\
\mu_{s} &=& \mu_{u}+\mu_{e} .
\end{eqnarray}
The electric charge neutrality condition is given by
\begin{equation}
\frac{2}{3}n_{u}-\frac{1}{3}n_{d}-\frac{1}{3}n_{s}-n_{e}=0.
\end{equation}
%The total baryon number density follows from
%\begin{equation}
%n_{b}=\frac{1}{3}(n_{u}+n_{s}+n_{d}) .
%\end{equation}
The energy density is given by
\begin{equation}
\epsilon_{Q}=\sum_{i=u,d,s,e}(\Omega_{i}+\mu_{i}n_{i})+B,
\label{eq:epsQ}
\end{equation}
and the corresponding pressure is obtained from
\begin{equation}
p_{Q}=-\sum_{i=u,d,s,e}\Omega_{i}-B ,
\label{eq:pQ}
\end{equation}
where $B$ denotes the bag constant.

MDM is a stable and self-interacting DM candidate that emerges from the parity symmetric extension of the Standard Model of particles \cite{Foot1991}. The idea of the possible existence of MDM can be traced back to 1956, when Lee and Yang proposed that the weak interactions is not parity symmetric \cite{Lee1956}, they also pointed  that even if the interactions of the known particle were to violate parity, the symmetry could be restored if a set of mirror particles existed. For details about MDM, see Refs. \cite{Blinnikov1982,Blinnikov1983,Khlopov1991,Foot2004,Berezhiani2004,BEREZHIANI2005,Okun2007,Ciarcelluti2010,Foot2014}.

In the minimal parity-symmetric extension of the standard model \cite{Foot1991,Pavsic1974}, the group structure is $G\otimes G$, where G is the gauge group of the standard model. In this model the two sectors are described by the same lagrangians, but where ordinary particles have left-handed interactions, mirror particles have right-handed interactions. Thus, the microphysics of MDM is the same as that of ordinary matter, and we will use the same EOS for SQM and MDM.

Following \cite{Sandin2009,Ciarcelluti2011}, we neglect the direct interaction between MDM and the ordinary matter, although these two sectors interact through the gravitational interaction. Except for gravity, MDM could interact with ordinary matter via so-called kinetic mixing of gauge bosons, or via unknown fields that carry both ordinary and mirror charges. However, if such interactions exist, they must be weak and can be well ignored in our study \cite{Sandin2009}.

\section{The structure and tidal deformability of SSs with a MDM core }\label{struc}
In the following, we use geometrized units $G=c=1$, and use the subscript $Q$ for SQM and $D$ for MDM.

To study the properties of SSs with a MDM core, we employ a two-fluid formalism where SQM and MDM sectors do not interact directly. However, these two sectors interact through the gravitational interaction in this formalism.

In the two-fluid formalism, the Tolman-Oppenheimer-Volkoff (TOV) equations are \citep[e.g.,][]{Das2020a,Ciancarella2020,Sandin2009,Ciarcelluti2011}
\begin{equation}
\frac{dm(r)}{dr}=4\pi \epsilon(r) r^{2},
\end{equation}

\begin{equation}
\frac{dp_{Q}(r)}{dr}=-\frac{[m(r)+4\pi r^{3}p(r)][\epsilon_{Q}(r)+p_{Q}(r)]}{r[r-2m(r)]},
\end{equation}

\begin{equation}
\frac{dp_{D}(r)}{dr}=-\frac{[m(r)+4\pi r^{3}p(r)][\epsilon_{D}(r)+p_{D}(r)]}{r[r-2m(r)]},
\end{equation}
where,
\begin{eqnarray}
\label{newp}
\epsilon(r) &=& \epsilon_{Q}(r)+\epsilon_{D}(r),\\
p(r) &=& p_{Q}(r)+p_{D}(r),
\label{neweps}
\end{eqnarray}

The dimensionless tidal deformability is defined as $\Lambda\equiv \lambda/M^{5}$, where $\lambda$ denotes the tidal deformability
parameter, which can be expressed in terms of the dimensionless tidal Love number $k_{2}$ as
$\lambda=\frac{2}{3}k_{2}R^{5}$
\citep{Hinderer2008,Flanagan2008,Damour2009,Hinderer2010}. Thus,
one has
\begin{equation}
\Lambda=\frac{2}{3}k_{2}\beta^{-5},
\end{equation}
where $\beta$ is compactness of the star, and it is defined as $\beta\equiv M/R$.

In the two-fluid formalism, the tidal Love number $k_{2}$ can be calculated using the expression
\citep{Lattimer2016}
\begin{equation}
k_{2}=\frac{8}{5}\frac{\beta^{5}z}{F},
\end{equation}
with
\begin{eqnarray}
z\equiv(1-2\beta)^{2}[2-y_{R}+2\beta(y_{R}-1)]
\label{z}
\end{eqnarray}
and
\begin{eqnarray}
\nonumber
F&\equiv&6\beta(2-y_{R})+6\beta^{2}(5y_{R}-8)+4\beta^{3}(13-11y_{R}) \\
&&  +4\beta^{4}(3y_{R}-2)+8\beta^{5}(1+y_{R})+3z\textrm{ln}(1-2\beta).~~~~
\label{f}
\end{eqnarray}

In Eqs.\ (\ref{z}) and (\ref{f}), $y_{R}\equiv y(R)-4\pi R^{3}\epsilon_{s}/M$, where $y(R)$ is the value of $y(r)$ at the surface of the star, and the second term of right hand side exists because there is a nonzero energy density $\epsilon_{s}$ just inside the surface of SSs \cite{Postnikov2010}. The quantity $y(r)$ satisfies the differential equation
\begin{equation}
\frac{dy(r)}{dr}=-\frac{y(r)^{2}}{r}-\frac{y(r)-6}{r-2m(r)}-rQ(r),
\label{yr}
\end{equation}
with \citep[e.g.,][]{Das2020a,Ciancarella2020}
\begin{eqnarray}
\nonumber Q(r)&\equiv&\frac{4\pi r}{r-2m(r)}
       \bigg[[5-y(r)]\epsilon(r)+[9+y(r)]p(r)
	\\ \nonumber &&+\frac{\epsilon_{Q}(r)+p_{Q}(r)}{\partial p_{Q}(r)/\partial \epsilon_{Q}(r)}+\frac{\epsilon_{D}(r)+p_{D}(r)}{\partial p_{D}(r)/\partial \epsilon_{D}(r)} \bigg]
    \\ &&-4\bigg[\frac{m(r)+4\pi r^{3}p(r)}{r[r-2m(r)]}\bigg]^{2} ,
\label{Qr}
\end{eqnarray}

For a given EOS, Eq.\ (\ref{yr}) can be calculated together with the TOV equations with the boundary conditions $y(0)=2$, $p(R)=0$, $m(0)=0$ for a given pressure at the center of the star $p(0)$. Note that there is another energy density jump $\epsilon_{sD}$ at the surface of the MDM core. Therefore, a correction of  $-4\pi R_D^{3}\epsilon_{sD}/M(R_D)$ is added to $y(R_D)$, where $R_D$ is the radius of the MDM core.

\section{results and discussions}\label{results}

\begin{figure*}
\centering
\resizebox{\hsize}{!}{\includegraphics{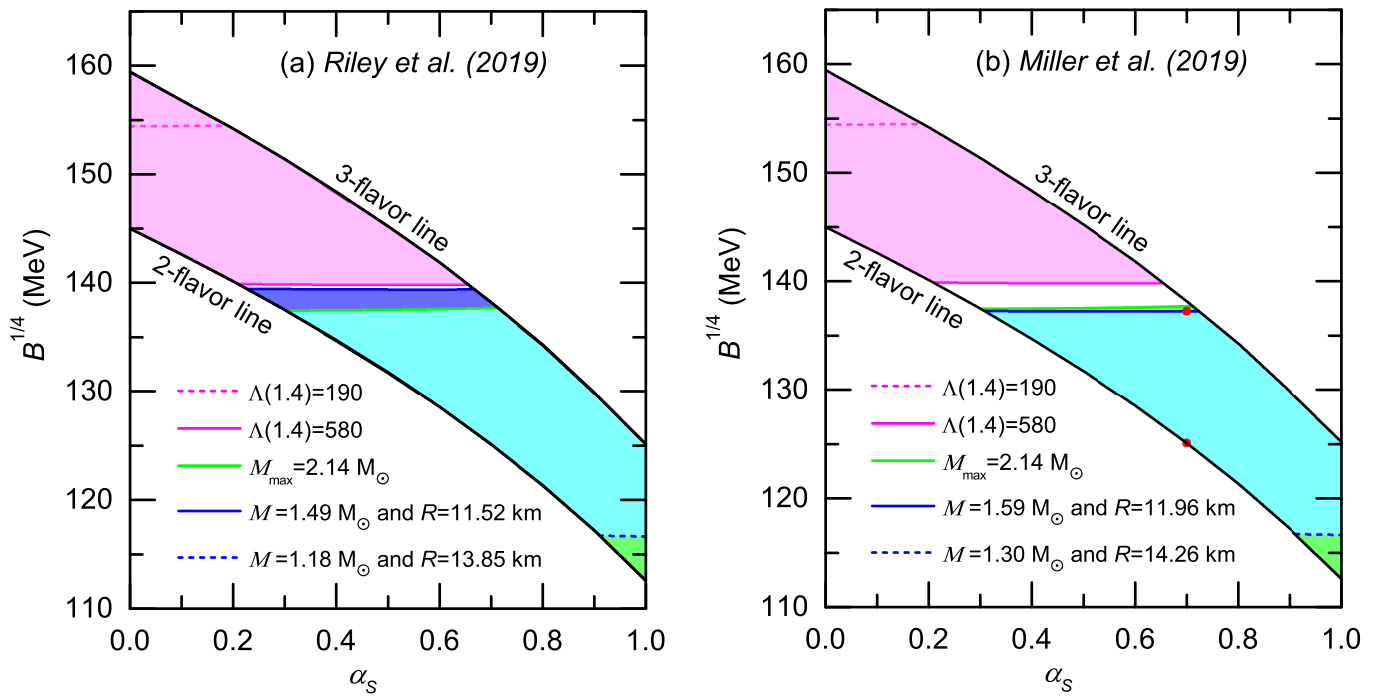}}
         \caption{(Color online) The constraints to the parameters of the EOS of SQM, namely, $B^{1/4}$ and $\alpha_{S}$ for $m_{s}=93$ MeV for the case of SSs without a MDM core. The magenta-shadowed area corresponds to the allowed parameter space according to the dimensionless tidal deformability of a $1.4\, M_{\odot}$ star of GW170817 ($\Lambda(1.4)=190 _{-120}^{+390}$). The cyan-shadowed area marks the parameter space which satisfy both $M_{\rm max} \geq 2.14\, M_{\odot}$ and the observational data of NICER (Fig.1a for the
         analysis by Riley et al.\ \cite{Riley2019} and Fig.1b for Miller et al.\ \cite{Miller2019}). The blue regions merely satisfy $M_{\rm max} \geq 2.14\, M_{\odot}$, and the green regions merely satisfy the observational data of NICER. The red dots in Fig.1b mark the two parameter sets (125.1,0.7) and (137.3, 0.7).
          }
   \label{fig1}
\end{figure*}

We investigate the allowed parameter space of the standard MIT bag model according to the following five constraints \citep[e.g.,][]
{Yang2020,Yang2021,Schaab1997,Weissenborn2011,Pi2015,Zhou2018}:

First, the existence of SSs is based on the idea that the presence of strange quarks lowers the energy per baryon of a mixture of $u$, $d$
and $s$ quarks in beta equilibrium below the energy of the most stable atomic nucleus, $^{56}$Fe ($E/A\sim 930$ MeV) \cite{Witten1984}. This
constraint results in the 3-flavor lines shown in Fig.\ \ref{fig1}.

The second constraint is given by assuming that non-strange quark matter (i.e., two-flavor quark matter made of only $u$ and $d$ quarks)
in bulk has an energy per baryon higher than the one of $^{56}$Fe, plus a 4 MeV correction coming from surface effects
\citep{Farhi1984,Madsen1999,Zhou2018}.  By imposing $E/A\geq 934$ MeV on non-strange quark matter, one ensures that atomic nuclei do not
dissolve into their constituent quarks. This leads to the 2-flavor lines in Fig.\ \ref{fig1}.  The areas between the 3-flavor lines and the 2-flavor
lines in Fig.\ \ref{fig1} show the allowed $B^{1/4}$--$\alpha_{S}$ parameter regions where the first and the second constraints described just above are fulfilled.

The above two constraints are from nuclear structure. As we discuss the constraints from astrophysics in the following, these two constraints must be fulfilled, which means that we are only interested in the region between the 3-flavor line and the 2-flavor line.

The third constraint is that the maximum mass of SSs must be greater than the mass of PSR J0740+6620, $M_{\rm max} \geq 2.14\,
M_{\odot}$. By employing this constraint, the allowed parameter space is limited to the region below the green line in Fig.\ \ref{fig1}.

The fourth constraint is form the observational data of NICER. The NICER observations of the isolated pulsar PSR J0030+0451 produced two independent measurements of the pulsar's mass and equatorial radius: $M = 1.34_{-0.16}^{+0.15}\, M_{\odot}$ and $R_{\rm eq} = 12.71_{-1.19}^{+1.14}$ km \citep{Riley2019}, and $M =1.44_{-0.14}^{+0.15}\, M_{\odot}$ and $R_{\rm eq}=13.02_{-1.06}^{+1.24}$ km \citep{Miller2019}. In Fig.\ \ref{fig1}, these
data on the $M$--$R$ plane is translated into the $B^{1/4}$--$\alpha_{S}$ space, which lead to the region between the blue solid and blue dashed lines.

The cyan-shadowed area Fig.\ \ref{fig1} marks the parameter space which satisfy all the four constraints mentioned above simultaneously.

The last constraint follows from the tidal deformability observation of GW170817, $\Lambda(1.4)=190 _{-120}^{+390}$, where $\Lambda(1.4)$ is the dimensionless tidal deformability of a $1.4\, M_{\odot}$ star. The parameter space satisfies this constraint corresponds to the magenta-shadowed area in Fig.\ \ref{fig1}.

Here we want to stress that Fig.\ \ref{fig1} is plotted for SSs without a MDM core. As can be seen from Fig.\ \ref{fig1}, the cyan-shadowed area does not overlap with the magenta-shadowed area, which means that SSs without a MDM core can not agree with the observations of PSR J0740+6620, PSR J0030+0451 and GW170817 simultaneously.

\begin{figure*}
\centering
\resizebox{\hsize}{!}{\includegraphics{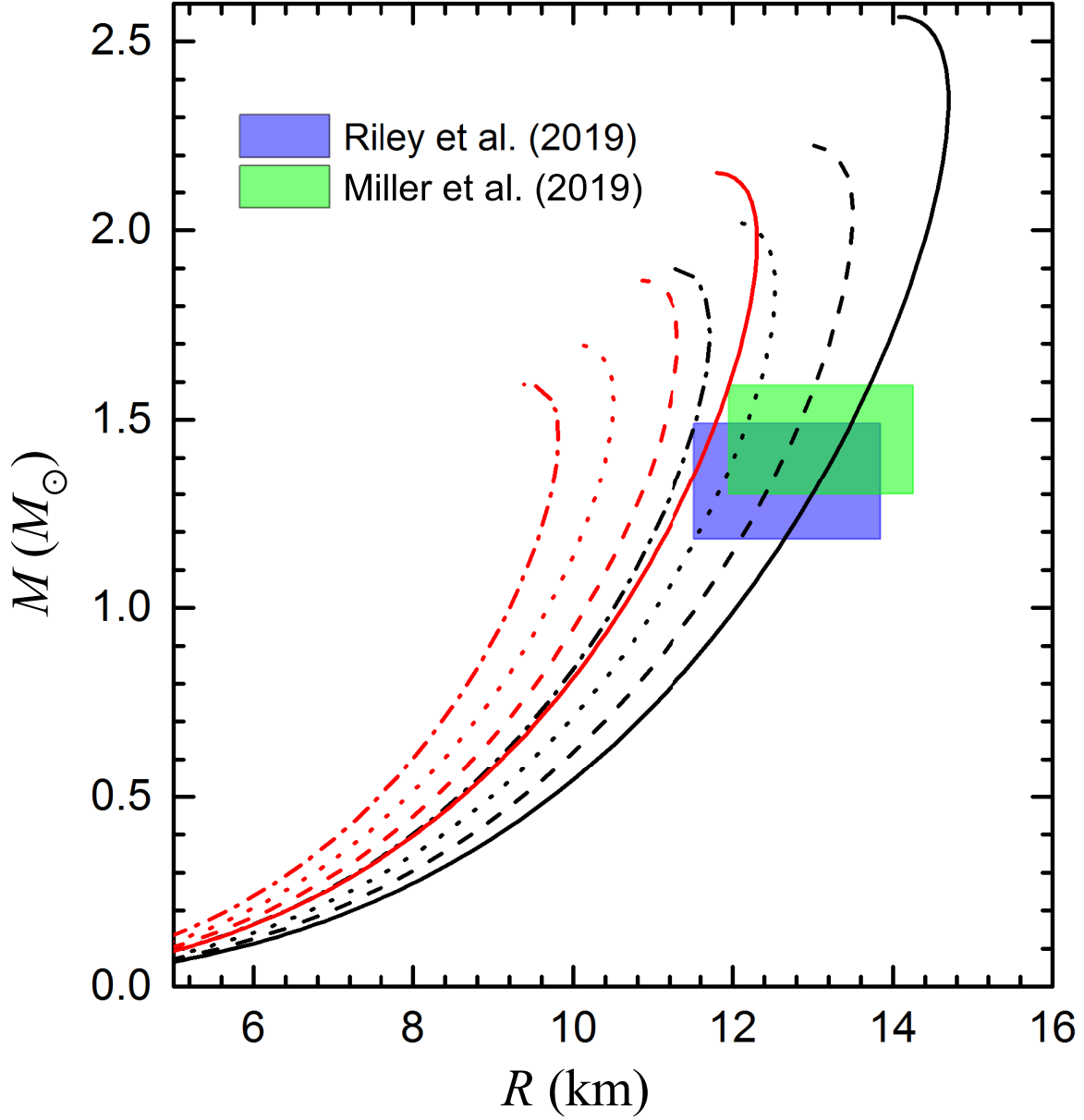}}
         \caption{(Color online) The mass-radius relation of SSs for $m_{s}=93$ MeV and $\alpha_{S}=0.7$ with various values of the mass fraction of MDM ($f_{D}$). The black lines are for $B^{1/4}=125.1$ MeV, and the red lines are for $B^{1/4}=137.3$ MeV. The solid, dashed, dotted, dash-dotted lines are for $f_{D}=0$\%, 10\%, 20\%, and 30\%, respectively. The blue and green regions show the mass and radius estimates of PSR J0030+0451 derived from NICER data by Riley et al. \cite{Riley2019} ($R = 12.71_{-1.19}^{+1.14}$ km, $M = 1.34_{-0.16}^{+0.15}\, M_{\odot}$) and Miller et al. \cite{Miller2019} ($R = 13.02_{-1.06}^{+1.24}$ km, $M=1.44_{-0.14}^{+0.15}\, M_{\odot}$).}
   \label{fig2}
\end{figure*}

Fig.\ \ref{fig2} shows the mass-radius relation of SSs for $m_{s}=93$ MeV and $\alpha_{S}=0.7$ with various values of $f_{D}$. The mass fraction of MDM ($f_{D}$) is defined by $f_{D}=M_{D}/M$, where $M$ is the total mass of the star and $M_{D}$ is the mass of the MDM core. As can be seen in Fig.\ \ref{fig2}, both the maximum mass and the radius of the $1.4\, M_{\odot}$ star decrease with the increasing of $f_{D}$.

\begin{figure*}
\centering
\resizebox{\hsize}{!}{\includegraphics{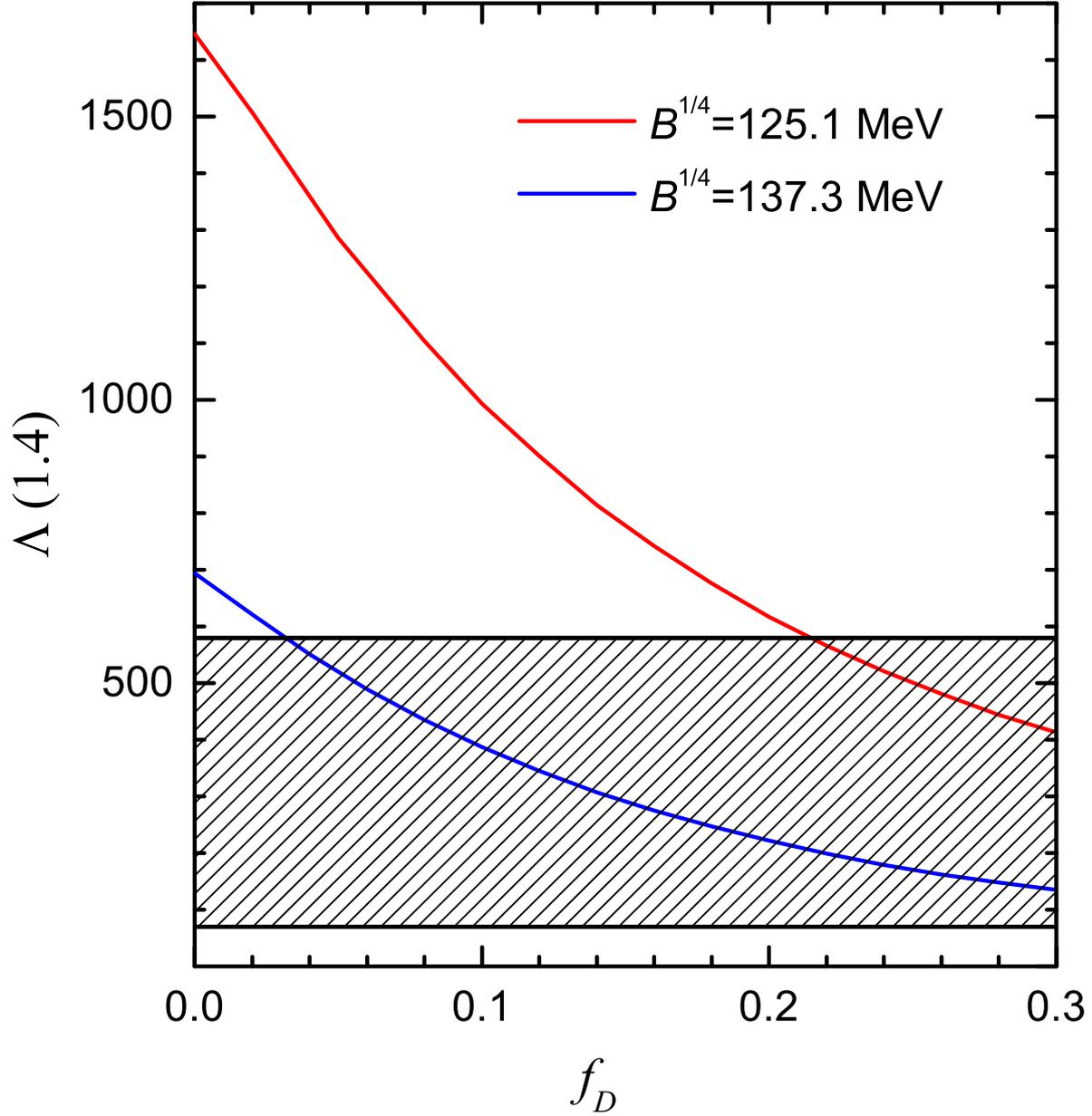}}
         \caption{(Color online) Relation between the dimensionless tidal deformability of a $1.4\, M_{\odot}$ star ($\Lambda(1.4)$) and the mass fraction of MDM ($f_{D}$) for $m_{s}=93$ MeV and $\alpha_{S}=0.7$. The shaded region corresponds to $70<\Lambda(1.4)<580$, which is given by the observation of GW170817.
          }
   \label{fig3}
\end{figure*}

Fig.\ \ref{fig3} shows the relation between the dimensionless tidal deformability of a $1.4\, M_{\odot}$ star ($\Lambda(1.4)$) and the mass fraction of MDM ($f_{D}$) for $m_{s}=93$ MeV and $\alpha_{S}=0.7$. One can easily find that $\Lambda(1.4)$ decrease with the increasing of $f_{D}$, and it is bigger for smaller value of $B^{1/4}$.

\begin{figure*}
\centering
\resizebox{\hsize}{!}{\includegraphics{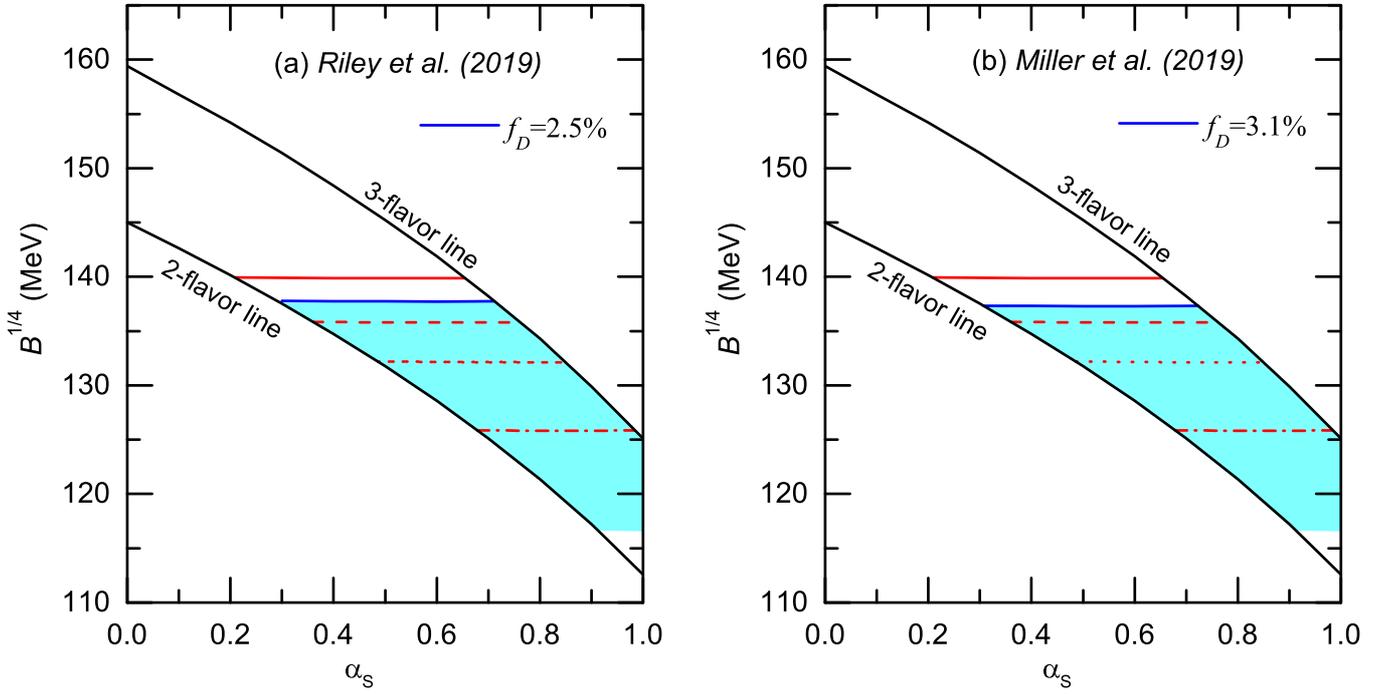}}
         \caption{(Color online) The constraints to the parameters of the EOS of SQM, namely, $B^{1/4}$ and $\alpha_{S}$ for $m_{s}=93$ MeV for SSs with various values of the mass fraction of MDM ($f_{D}$). Similar to Fig.1, the cyan-shadowed area marks the parameter space which satisfy both $M_{\rm max} \geq 2.14\, M_{\odot}$ and the observational data of NICER (Fig.4a for the analysis by Riley et al.\ \cite{Riley2019} and Fig.4b for Miller et al.\ \cite{Miller2019}) for SSs without a MDM core (in other words, for $f_{D}=0$\%). The red lines are for $\Lambda(1.4)=580$, with solid, dashed, dotted, dash-dotted lines for $f_{D}=0$\%, 5\%, 10\%, and 20\%, respectively. The blue line is for $f_{D}=2.5$\% in Fig.4a, and is for $f_{D}=3.1$\% in Fig.4b.
         }
   \label{fig4}
\end{figure*}

Fig.\ \ref{fig4} shows the constraints to the parameters of the EOS of SQM for $m_{s}=93$ MeV for SSs with various values of $f_{D}$. Note that the cyan-shadowed areas are the same as these in Fig.\ \ref{fig1}, which are for the case of SSs without a MDM core (in other words, for $f_{D}=0$\%) and satisfy both $M_{\rm max} \geq 2.14\, M_{\odot}$ and the observational data of NICER. We can see that the parameter space region which satisfy $\Lambda(1.4) < 580$ (the region above the $\Lambda(1.4)=580$ line) shifts downward as $f_{D}$ increases and it begins to overlap with the cyan-shadowed area for the value of $f_{D}=2.5$\% for the analysis of NICER data by Riley et al.\ \cite{Riley2019} (Fig.4a), and $f_{D}=3.1$\% for Miller et al.\ \cite{Miller2019} (Fig.4b). Thus, assuming PSR J0740+6620 and PSR J0030+0451 are SSs without a MDM core, there exists allowed parameter space for which SSs agree with the observations of PSR J0740+6620, PSR J0030+0451 and GW170817 simultaneously in the case that SSs in GW170817 have a MDM core with $f_{D}>2.5$\% (for the case of Riley et al., and $f_{D}>3.1$\% for the case of Miller et al.).

\section{Summary and conclusions}\label{Summary}
In this Letter, we study the structure and the tidal deformability of SSs with a MDM core and explain the observations of PSR J0740+6620, PSRJ0030+0451 and GW170817 simultaneously. As realized by \cite{Sandin2009,Ciarcelluti2011}, The mass fraction of MDM ($f_{D}$) of each SS could be different, which depends on the individual history. We show that all the above observations could be explained simultaneously if one assumes that PSR J0740+6620 and PSR J0030+0451 are SSs without a MDM core, and SSs in GW170817 have a MDM core with $f_{D}>2.5$\% (for the case of Riley et al., and $f_{D}>3.1$\% for the case of Miller et al.). In fact, it is easy to deduce that in order to fulfill all these observations, it is not necessary to assume that PSR J0740+6620 and PSR J0030+0451 are SSs without a MDM core. PSR J0740+6620 and PSR J0030+0451 could be SSs with a MDM core, but SSs in GW170817 should have a larger MDM core than them in this case. As a conclusion, to explain of all the observations, SSs in GW170817 should have a MDM core.

Finally, as pointed out by Ciarcelluti and Sandin \cite{Ciarcelluti2011}, although our results are calculated for MDM, they are qualitatively valid for other kinds of dark matter that could form stable cores inside SSs. Therefore, our study leads to the result that for the standard MIT bag model, the observations of PSR J0740+6620, PSR J0030+0451 and GW170817 confirm the existence of a dark-matter core inside SSs.

%\cite{Yang2020,Yang2021,Ciancarella2020,Ciarcelluti2011,Sandin2009,Abbott2017,Abbott2018,Riley2019,Miller2019,Cromartie2020,Zyla2020}.
%dark matter admixed strange stars:\cite{Mukhopadhyay2016,Panotopoulos2017}
%dark matter admixed neutron stars:\cite{Leung2011,Li2012,Li2012a,Xiang2014,Mukhopadhyay2017,Ellis2018,Rezaei2018,Wang2019,Deliyergiyev2019,Ivanytskyi2020,Das2020a}
%dark matter admixed neutron stars,including tidal deformability:
%dark matter admixed neutron stars, postmerger GW gravitational signals from neutron star mergers:\cite{Ellis2018a,Bezares2019,Bauswein2020}
%The 1991 paper by Foot et al, 336 citing on ADS now: \cite{Foot1991}
%The 1956 paper of Lee and Yang: \cite{Lee1956}
%MDM review references:
%\cite{Foot2004,Berezhiani2004,BEREZHIANI2005,Okun2007,Ciarcelluti2010,Foot2014}
%%%%%%%%%%%%%%%%%%%%%%%%%%%%%%%%%%%%%
\section*{Acknowledgments}
%%%%%%%%%%%%%%%%%%%%%%%%%%%%%%%%%%%%%%%%%%%%%%%%%%%%%%%%%%
This work is supported by the Scientific Research Program of the National Natural Science Foundation of China (NSFC, grant Nos.\ 12033001, 11773011, and 11447012), and National SKA Program of China No. 2020SKA0120300.

\section*{References}

\bibliography{mbm}

\begin{thebibliography}{10}
\expandafter\ifx\csname url\endcsname\relax
  \def\url#1{\texttt{#1}}\fi
\expandafter\ifx\csname urlprefix\endcsname\relax\def\urlprefix{URL }\fi
\expandafter\ifx\csname href\endcsname\relax
  \def\href#1#2{#2} \def\path#1{#1}\fi

\bibitem{Itoh1970}
N.~Itoh, Hydrostatic equilibrium of hypothetical quark stars, Prog. Theor.
  Phys. 44 (1970) 291--292.
\newblock \href {http://dx.doi.org/10.1143/PTP.44.291}
  {\path{doi:10.1143/PTP.44.291}}.

\bibitem{Bodmer1971}
A.~R. Bodmer, Collapsed nuclei, Phys. Rev. D 4 (1971) 1601--1606.
\newblock \href {http://dx.doi.org/10.1103/PhysRevD.4.1601}
  {\path{doi:10.1103/PhysRevD.4.1601}}.

\bibitem{Witten1984}
E.~Witten, Cosmic separation of phases, Phys. Rev. D 30 (1984) 272--285.
\newblock \href {http://dx.doi.org/10.1103/PhysRevD.30.272}
  {\path{doi:10.1103/PhysRevD.30.272}}.

\bibitem{Terazawa1989}
H.~Terazawa, Super-hypernuclei in the quark-shell model, J. Phys. Soc. Japan 58
  (1989) 3555.
\newblock \href {http://dx.doi.org/10.1143/JPSJ.58.3555}
  {\path{doi:10.1143/JPSJ.58.3555}}.

\bibitem{Farhi1984}
E.~Farhi, R.~L. Jaffe, Strange matter, Phys. Rev. D 30 (1984) 2379--2390.
\newblock \href {http://dx.doi.org/10.1103/PhysRevD.30.2379}
  {\path{doi:10.1103/PhysRevD.30.2379}}.

\bibitem{Alcock1986}
C.~Alcock, E.~Farhi, A.~Olinto, Strange stars, Astrophys. J. 310 (1986) 261.
\newblock \href {http://dx.doi.org/10.1086/164679} {\path{doi:10.1086/164679}}.

\bibitem{Haensel1986}
P.~Haensel, J.~L. Zdunik, R.~Schaefer, Strange quark stars, Astron. Astrophys.
  160 (1986) 121--128.

\bibitem{Alcock1988}
C.~Alcock, A.~Olinto, Exotic phases of hadronic matter and their astrophysical
  application, Ann. Rev. Nucl. Part. Sci. 38 (1988) 161--184.
\newblock \href {http://dx.doi.org/10.1146/annurev.ns.38.120188.001113}
  {\path{doi:10.1146/annurev.ns.38.120188.001113}}.

\bibitem{Madsen1999}
J.~Madsen, Physics and astrophysics of strange quark matter, Lect.Notes Phys.
  516 (1999) 162--203.
\newblock \href {http://dx.doi.org/10.1007/BFb0107314}
  {\path{doi:10.1007/BFb0107314}}.

\bibitem{Weber2005}
F.~Weber, Strange quark matter and compact stars, Prog. Part. Nucl. Phys. 54
  (2005) 193--288.
\newblock \href {http://dx.doi.org/10.1016/j.ppnp.2004.07.001}
  {\path{doi:10.1016/j.ppnp.2004.07.001}}.

\bibitem{Yang2020}
S.-H. Yang, C.-M. Pi, X.-P. Zheng, F.~Weber, Non-newtonian gravity in strange
  quark stars and constraints from the observations of {PSR} {J}0740-6620 and
  {GW}170817, Astrophys. J. 902 (2020) 32.
\newblock \href {http://dx.doi.org/10.3847/1538-4357/abb365}
  {\path{doi:10.3847/1538-4357/abb365}}.

\bibitem{Yang2021}
S.-H. Yang, C.-M. Pi, X.-P. Zheng, F.~Weber, Constraints from compact star
  observations on non-{N}ewtonian gravity in strange stars based on a density
  dependent quark mass model, Phys. Rev. D 103 (2021) 043012.
\newblock \href {http://dx.doi.org/10.1103/PhysRevD.103.043012}
  {\path{doi:10.1103/PhysRevD.103.043012}}.

\bibitem{Cromartie2020}
H.~Cromartie, E.~Fonseca, S.~M. Ransom, P.~B. Demorest, Z.~Arzoumanian, et~al.,
  Relativistic shapiro delay measurements of an extremely massive millisecond
  pulsar, Nat. Astron. 4 (2020) 72--76.
\newblock \href {http://dx.doi.org/10 .1038 /s41550 -019 -0880 -2}
  {\path{doi:10 .1038 /s41550 -019 -0880 -2}}.

\bibitem{Abbott2017}
B.~P. Abbott, R.~Abbott, T.~D. Abbott, F.~Acernese, K.~Ackley, et~al.,
  {GW}170817: observation of gravitational waves from a binary neutron star
  inspiral, Phys. Rev. Lett. 119 (2017) 161101.
\newblock \href {http://dx.doi.org/10 .1103 /PhysRevLett .119 .161101}
  {\path{doi:10 .1103 /PhysRevLett .119 .161101}}.

\bibitem{Abbott2018}
B.~P. Abbott, R.~Abbott, T.~D. Abbott, F.~Acernese, K.~Ackley, et~al.,
  {GW}170817: measurements of neutron star radii and equation of state, Phys.
  Rev. Lett. 121 (2018) 161101.
\newblock \href {http://dx.doi.org/10.1103/physrevlett.121.161101}
  {\path{doi:10.1103/physrevlett.121.161101}}.

\bibitem{Spergel2000}
D.~N. Spergel, P.~J. Steinhardt, Observational evidence for self-interacting
  cold dark matter, Phys. Rev. Lett. 84 (2000) 3760--3763.
\newblock \href {http://dx.doi.org/10.1103/PhysRevLett.84.3760}
  {\path{doi:10.1103/PhysRevLett.84.3760}}.

\bibitem{Tulin2018}
S.~Tulin, H.-B. Yu, Dark matter self-interactions and small scale structure,
  Phys. Rep. 730 (2018) 1--57.
\newblock \href {http://dx.doi.org/10.1016/j.physrep.2017.11.004}
  {\path{doi:10.1016/j.physrep.2017.11.004}}.

\bibitem{Bertone2018}
G.~Bertone, T.~M.~P. Tait, A new era in the search for dark matter, Nature 562
  (2018) 51--56.
\newblock \href {http://dx.doi.org/10.1038/s41586-018-0542-z}
  {\path{doi:10.1038/s41586-018-0542-z}}.

\bibitem{Leung2011}
S.-C. Leung, M.-C. Chu, L.-M. Lin, Dark-matter admixed neutron stars, Phys.
  Rev. D 84 (2011) 107301.
\newblock \href {http://dx.doi.org/10.1103/PhysRevD.84.107301}
  {\path{doi:10.1103/PhysRevD.84.107301}}.

\bibitem{Li2012}
X.~Y. Li, F.~Y. Wang, K.~S. Cheng, Gravitational effects of condensate dark
  matter on compact stellar objects, J. Cosmol. Astropart. Phys. 1210 (2012)
  31.
\newblock \href {http://dx.doi.org/10.1088/1475-7516/2012/10/031}
  {\path{doi:10.1088/1475-7516/2012/10/031}}.

\bibitem{Li2012a}
A.~Li, F.~Huang, R.-X. Xu, Too massive neutron stars: The role of dark matter?,
  Astropart. Phys. 37 (2012) 70--74.
\newblock \href {http://dx.doi.org/10.1016/j.astropartphys.2012.07.006}
  {\path{doi:10.1016/j.astropartphys.2012.07.006}}.

\bibitem{Xiang2014}
Q.-F. Xiang, W.-Z. Jiang, D.-R. Zhang, R.-Y. Yang, Effects of fermionic dark
  matter on properties of neutron stars, Phys. Rev. C 89 (2014) 025803.
\newblock \href {http://dx.doi.org/10.1103/PhysRevC.89.025803}
  {\path{doi:10.1103/PhysRevC.89.025803}}.

\bibitem{Mukhopadhyay2017}
S.~Mukhopadhyay, D.~Atta, K.~Imam, D.~N. Basu, C.~Samanta, Compact bifluid
  hybrid stars: hadronic matter mixed with self-interacting fermionic
  asymmetric dark matter, Eur. Phys. J. C 77 (2017) 440.
\newblock \href {http://dx.doi.org/10.1140/epjc/s10052-017-5006-3}
  {\path{doi:10.1140/epjc/s10052-017-5006-3}}.

\bibitem{Ellis2018}
J.~Ellis, G.~Hutsi, K.~Kannike, L.~Marzola, M.~Raidal, V.~Vaskonen, Dark matter
  effects on neutron star properties, Phys. Rev. D 97 (2018) 123007.
\newblock \href {http://dx.doi.org/10.1103/PhysRevD.97.123007}
  {\path{doi:10.1103/PhysRevD.97.123007}}.

\bibitem{Ellis2018a}
J.~Ellis, A.~Hektor, G.~Hutsi, K.~Kannike, L.~Marzola, M.~Raidal, V.~Vaskonen,
  Search for dark matter effects on gravitational signals from neutron star
  mergers, Phys. Lett. B 781 (2018) 607--610.
\newblock \href {http://dx.doi.org/10.1016/j.physletb.2018.04.048}
  {\path{doi:10.1016/j.physletb.2018.04.048}}.

\bibitem{Rezaei2018}
Z.~Rezaei, Neutron stars with spin polarized self-interacting dark matter,
  Astropart. Phys. 101 (2018) 1--7.
\newblock \href {http://dx.doi.org/10.1016/j.astropartphys.2018.03.002}
  {\path{doi:10.1016/j.astropartphys.2018.03.002}}.

\bibitem{Wang2019}
X.~D. Wang, B.~Qi, G.~L. Yang, N.~B. Zhang, S.~Y. Wang, Possible maximum mass
  of dark matter existing in compact stars based on the self-interacting
  fermionic model, Int. J. Mod. Phys. D 28 (2019) 1950148.
\newblock \href {http://dx.doi.org/10.1142/S0218271819501487}
  {\path{doi:10.1142/S0218271819501487}}.

\bibitem{Deliyergiyev2019}
M.~Deliyergiyev, A.~D. Popolo, L.~Tolos, M.~L. Delliou, X.~Lee, F.~Burgio, Dark
  compact objects: An extensive overview, Phys. Rev. D 99 (2019) 063015.
\newblock \href {http://dx.doi.org/10.1103/PhysRevD.99.063015}
  {\path{doi:10.1103/PhysRevD.99.063015}}.

\bibitem{Bezares2019}
M.~Bezares, D.~Vigano, C.~Palenzuela, Gravitational wave signatures of dark
  matter cores in binary neutron star mergers by using numerical simulations,
  Phys. Rev. D 100 (2019) 044049.
\newblock \href {http://dx.doi.org/10.1103/PhysRevD.100.044049}
  {\path{doi:10.1103/PhysRevD.100.044049}}.

\bibitem{Ivanytskyi2020}
O.~Ivanytskyi, V.~Sagun, I.~Lopes, Neutron stars: New constraints on asymmetric
  dark matter, Phys. Rev. D 102 (2020) 063028.
\newblock \href {http://dx.doi.org/10.1103/PhysRevD.102.063028}
  {\path{doi:10.1103/PhysRevD.102.063028}}.

\bibitem{Bauswein2020}
A.~Bauswein, G.~Guo, J.-H. Lien, Y.-H. Lin, M.-R. Wu, Compact dark objects in
  neutron star mergers.\href {http://arxiv.org/abs/2012.11908}
  {\path{arXiv:2012.11908}}.

\bibitem{Das2020a}
A.~Das, T.~Malik, A.~C. Nayak, Dark matter admixed neutron star properties in
  the light of gravitational wave observations: a two fluid approach.\href
  {http://arxiv.org/abs/2011.01318} {\path{arXiv:2011.01318}}.

\bibitem{Ciancarella2020}
R.~Ciancarella, F.~Pannarale, A.~Addazi, A.~Marciano, Constraining mirror dark
  matter inside neutron stars.\href {http://arxiv.org/abs/2010.12904}
  {\path{arXiv:2010.12904}}.

\bibitem{Mukhopadhyay2016}
P.~Mukhopadhyay, J.~Schaffner-Bielich, Quark stars admixed with dark matter,
  Phys. Rev. D 93 (2016) 083009.
\newblock \href {http://dx.doi.org/10.1103/PhysRevD.93.083009}
  {\path{doi:10.1103/PhysRevD.93.083009}}.

\bibitem{Panotopoulos2017}
G.~Panotopoulos, I.~Lopes, Gravitational effects of condensed dark matter on
  strange stars, Phys. Rev. D 96 (2017) 023002.
\newblock \href {http://dx.doi.org/10.1103/PhysRevD.96.023002}
  {\path{doi:10.1103/PhysRevD.96.023002}}.

\bibitem{Sandin2009}
F.~Sandin, P.~Ciarcelluti, Effects of mirror dark matter on neutron stars,
  Astropart. Phys. 32 (2009) 278--284.
\newblock \href {http://dx.doi.org/10.1016/j.astropartphys.2009.09.005}
  {\path{doi:10.1016/j.astropartphys.2009.09.005}}.

\bibitem{Ciarcelluti2011}
P.~Ciarcelluti, F.~Sandin, Have neutron stars a dark matter core?, Phys. Lett.
  B 695 (2011) 19--21.
\newblock \href {http://dx.doi.org/10.1016/j.physletb.2010.11.021}
  {\path{doi:10.1016/j.physletb.2010.11.021}}.

\bibitem{Ozel2006}
F.~Ozel, Soft equations of state for neutron-star matter ruled out by {EXO}
  0748 - 676, Nature 441 (2006) 1115.
\newblock \href {http://dx.doi.org/10.1038/nature04858}
  {\path{doi:10.1038/nature04858}}.

\bibitem{Riley2019}
T.~E. Riley, A.~L. Watts, S.~Bogdanov, P.~S. Ray, R.~M. Ludlam, et~al., {A}
  {NICER} view of {PSR} {J}0030+0451: millisecond pulsar parameter estimation,
  Astrophys. J. Lett. 887 (2019) L21.
\newblock \href {http://dx.doi.org/10 .3847 /2041 -8213 /ab481c} {\path{doi:10
  .3847 /2041 -8213 /ab481c}}.

\bibitem{Miller2019}
M.~C. Miller, F.~K. Lamb, A.~J. Dittmann, S.~Bogdanov, Z.~Arzoumanian, et~al.,
  {PSR} {J}0030+0451 mass and radius from {NICER} data and implications for the
  properties of neutron star matter, Astrophys. J. Lett. 887 (2019) L24.
\newblock \href {http://dx.doi.org/10.3847/2041-8213/ab50c5}
  {\path{doi:10.3847/2041-8213/ab50c5}}.

\bibitem{Zyla2020}
P.~A. Zyla, R.~M. Barnett, J.~Beringer, O.~Dahl, D.~A. Dwyer, et~al., Review of
  particle physics, Prog. Theor. Exp. Phys. 2020 (2020) 083C01.
\newblock \href {http://dx.doi.org/10.1093/ptep/ptaa104}
  {\path{doi:10.1093/ptep/ptaa104}}.

\bibitem{Foot1991}
R.~Foot, H.~Lew, R.~R. Volkas, A model with fundamental improper spacetime
  symmetries, Phys. Lett. B 272 (1991) 67--70.
\newblock \href {http://dx.doi.org/10.1016/0370-2693(91)91013-L}
  {\path{doi:10.1016/0370-2693(91)91013-L}}.

\bibitem{Lee1956}
T.~D. Lee, C.~N. Yang, Question of parity conservation in weak interactions,
  Phys. Rev. 104 (1956) 254--258.
\newblock \href {http://dx.doi.org/10.1103/PhysRev.104.254}
  {\path{doi:10.1103/PhysRev.104.254}}.

\bibitem{Blinnikov1982}
S.~I. Blinnikov, M.~Y. Khlopov, On the possible signatures of mirror particles,
  Sov. J. Nucl. Phys. 36 (1982) 472--474.

\bibitem{Blinnikov1983}
S.~I. Blinnikov, M.~Y. Khlopov, Possible astronomical effects of mirror
  particles, Sov. Astron. 27 (1983) 371--375.

\bibitem{Khlopov1991}
M.~Y. Khlopov, G.~M. Beskin, N.~G. Bochkarev, L.~A. Pustilnik, S.~A. Pustilnik,
  Observational physics of the mirror world,, Sov. Astron. 35 (1991) 21--30.

\bibitem{Foot2004}
R.~Foot, Mirror matter-type dark matter, Int. J. Mod. Phys. D 13 (2004)
  2161--2192.
\newblock \href {http://dx.doi.org/10.1142/S0218271804006449}
  {\path{doi:10.1142/S0218271804006449}}.

\bibitem{Berezhiani2004}
Z.~Berezhiani, Mirror world and its cosmological consequences, Int. J. Mod.
  Phys. A 19 (2004) 3775--3806.
\newblock \href {http://dx.doi.org/10.1142/S0217751X04020075}
  {\path{doi:10.1142/S0217751X04020075}}.

\bibitem{BEREZHIANI2005}
Z.~Berezhiani, Through the looking-glass: {A}lice's adventures in mirror world,
  in: {F}rom {F}ields to {S}trings: {C}ircumnavigating {T}heoretical
  {P}hysics,{V}ol. 3, eds. {M}. {S}hifman et al. ({W}orld {S}cientific) (2005)
  2147--2195\href {http://dx.doi.org/10.1142/9789812775344_0055}
  {\path{doi:10.1142/9789812775344_0055}}.

\bibitem{Okun2007}
L.~B. Okun, Mirror particles and mirror matter: 50 years of speculation and
  searching, Phys. Usp. 50 (2007) 380--389.
\newblock \href {http://dx.doi.org/10.1070/pu2007v050n04abeh006227}
  {\path{doi:10.1070/pu2007v050n04abeh006227}}.

\bibitem{Ciarcelluti2010}
P.~Ciarcelluti, Cosmology with mirror dark matter, Int. J. Mod. Phys. D 19
  (2010) 2151--2230.
\newblock \href {http://dx.doi.org/10.1142/S0218271810018438}
  {\path{doi:10.1142/S0218271810018438}}.

\bibitem{Foot2014}
R.~Foot, Mirror dark matter: Cosmology, galaxy structure and direct detection,
  Int. J. Mod. Phys. A 29 (2014) 1430013.
\newblock \href {http://dx.doi.org/10.1142/S0217751X14300130}
  {\path{doi:10.1142/S0217751X14300130}}.

\bibitem{Pavsic1974}
M.~Pavsic, External inversion, internal inversion, and reflection invariance,
  Int. J. Theor. Phys. 9 (1974) 229--244.
\newblock \href {http://dx.doi.org/10.1007/BF01810695}
  {\path{doi:10.1007/BF01810695}}.

\bibitem{Hinderer2008}
T.~Hinderer, Tidal love numbers of neutron stars, Astrophys. J. 677 (2008)
  1216--1220.
\newblock \href {http://dx.doi.org/10.1086/533487} {\path{doi:10.1086/533487}}.

\bibitem{Flanagan2008}
E.~E. Flanagan, T.~Hinderer, Constraining neutron-star tidal love numbers with
  gravitational-wave detectors, Phys. Rev. D 77 (2008) 021502.
\newblock \href {http://dx.doi.org/10.1103/PhysRevD.77.021502}
  {\path{doi:10.1103/PhysRevD.77.021502}}.

\bibitem{Damour2009}
T.~Damour, A.~Nagar, Relativistic tidal properties of neutron stars, Phys. Rev.
  D 80 (2009) 084035.
\newblock \href {http://dx.doi.org/10.1103/PhysRevD.80.084035}
  {\path{doi:10.1103/PhysRevD.80.084035}}.

\bibitem{Hinderer2010}
T.~Hinderer, B.~D. Lackey, R.~N. Lang, J.~S. Read, Tidal deformability of
  neutron stars with realistic equations of state and their gravitational wave
  signatures in binary inspiral, Phys. Rev. D 81 (2010) 123016.
\newblock \href {http://dx.doi.org/10.1103/PhysRevD.81.123016}
  {\path{doi:10.1103/PhysRevD.81.123016}}.

\bibitem{Lattimer2016}
J.~M. Lattimer, M.~Prakash, The equation of state of hot, dense matter and
  neutron stars, Phys. Rep. 621 (2016) 127--164.
\newblock \href {http://dx.doi.org/10.1016/j.physrep.2015.12.005}
  {\path{doi:10.1016/j.physrep.2015.12.005}}.

\bibitem{Postnikov2010}
S.~Postnikov, M.~Prakash, J.~M. Lattimer, Tidal love numbers of neutron and
  self-bound quark stars, Phys. Rev. D 82 (2010) 024016.
\newblock \href {http://dx.doi.org/10.1103/PhysRevD.82.024016}
  {\path{doi:10.1103/PhysRevD.82.024016}}.

\bibitem{Schaab1997}
C.~Schaab, B.~Hermann, F.~Weber, M.~K. Weigel, Are strange stars
  distinguishable from neutron stars by their cooling behaviour?, J. Phys. G:
  Nucl. Part. Phys. 23 (1997) 2029--2037.
\newblock \href {http://dx.doi.org/10.1088/0954-3899/23/12/027}
  {\path{doi:10.1088/0954-3899/23/12/027}}.

\bibitem{Weissenborn2011}
S.~Weissenborn, I.~Sagert, G.~Pagliara, M.~Hempel, J.~Schaffner-Bielich, Quark
  matter in massive compact stars, Astrophys. J. Lett. 740 (2011) L14.
\newblock \href {http://dx.doi.org/10.1088/2041-8205/740/1/L14}
  {\path{doi:10.1088/2041-8205/740/1/L14}}.

\bibitem{Pi2015}
C.-M. Pi, S.-H. Yang, X.-P. Zheng, R-mode instability of strange stars and
  observations of neutron stars in {LMXB}s, Res. Astron. Astrophys. 15 (2015)
  871.
\newblock \href {http://dx.doi.org/10.1088/1674-4527/15/6/009}
  {\path{doi:10.1088/1674-4527/15/6/009}}.

\bibitem{Zhou2018}
E.-P. Zhou, X.~Zhou, A.~Li, Constraints on interquark interaction parameters
  with {GW}170817 in a binary strange star scenario, Phys. Rev. D 97 (2018)
  083015.
\newblock \href {http://dx.doi.org/10.1103/PhysRevD.97.083015}
  {\path{doi:10.1103/PhysRevD.97.083015}}.

\end{thebibliography}

\clearpage

\clearpage

\clearpage

\end{document}